\documentclass[aps,prb,twocolumn,showpacs]{revtex4}
\usepackage{amsmath}
\usepackage{graphicx,epsfig,psfrag}
\usepackage{amssymb}
\usepackage{hyperref}

\begin{document}
\title{Mechanism for large thermoelectric power in molecular 
quantum dots described by the negative-$U$ Anderson model} 
\author{S. Andergassen${}^1$}
\author{T. A. Costi${}^2$}
\author{V. Zlati\'c${}^{2,3,4}$}
\affiliation{
${}^1$Institut f\"{u}r Theorie der Statistischen Physik, 
RWTH Aachen University and JARA-Fundamentals of Information Technology, 
D-52056 Aachen, Germany\\
${}^2$Peter Gr\"{u}nberg Institut and Institute for Advanced Simulation,
Research Centre J\"ulich, D-52425 J\"ulich, Germany\\
${}^3$Institute of Physics, HR-10001 Zagreb, Croatia\\
${}^4$J. Stefan Institute, SI-1000 Ljubljana, Slovenia
}
\begin{abstract}
  We investigate with the aid of numerical renormalization group techniques 
  the thermoelectric properties of a molecular quantum dot 
  described by the negative-$U$ Anderson model. We show that the charge Kondo effect
  provides a mechanism for enhanced thermoelectric power via a correlation-induced 
  asymmetry in the spectral function close to the Fermi level. 
  We show that this effect results in a dramatic enhancement of the Kondo-induced peak 
  in the thermopower of negative-$U$ systems with Seebeck coefficients exceeding 
  50$\mu V/K$ over a wide range of gate voltages. 
\end{abstract}

\vspace{-0.8cm}
%71.27.+a
%72.15.Jf Thermoelectric effects
%72.10.Fk Scattering by point defects, dislocations, surfaces,
% and other imperfections (including Kondo effect)
%72.15.Qm Scattering mechanisms and Kondo effect
%72.10.-d Theory of electronic transport; scattering mechanisms
%73.63.Kv Quantum dots (transport)

\date{\today}
\maketitle
{\em Introduction.---} 
Thermoelectric devices currently use bulk materials, e.g., Si-Ge, 
PbTe or Bi$_2$Te$_3$ \cite{mahan.97,snyder.08,kanatzidis.10}. In the future, devices 
made of nanoscale objects, such as quantum dots or molecules, could offer 
alternatives, particularly for low-temperature applications, such as on-chip 
cooling of microprocessors or low-temperature refrigeration. 
Nanoscale objects have some potential advantages over
their bulk counterparts, for example, in scalability or in their high degree 
of tunability (e.g., via a gate voltage), allowing them to be operated at 
optimal thermoelectric efficiency. 
Molecular quantum dots, in particular, could be interesting to study, 
since a large variety of such systems could be fabricated and 
investigated for interesting thermoelectric properties \cite{reddy.07}. 

The description of electrical and thermal transport through quantum dots is, 
however, a challenging theoretical task. Electrons tunneling from the leads through
the quasi-localized levels of the dot typically experience a large 
Coulomb repulsion on the dot, giving rise to the spin Kondo effect \cite{hewson.97}. 
The latter profoundly affects transport, resulting, for example, in the 
lifting of Coulomb blockade at low temperatures for a wide range of gate voltages
and an enhanced conductance close to the unitary limit, $G\approx G_{0} = 2e^{2}/h$, for
symmetric coupling to the leads \cite{glazman.88,ng.88,goldhaber-gordon.98,
cronenwett.98,vanderwiel.00}. Recent experimental and theoretical work 
has also addressed the effects of Kondo correlations on the thermoelectric 
properties of such quantum dots \cite{scheibner.05,kim.02,costi.10}. However, the 
Kondo-induced enhancement of the thermopower at the Kondo temperature 
$T_{K}$  was found to be very small \cite{costi.10}, 
suggesting that the spin Kondo effect, in its simplest manifestation,
is ineffective for realizing efficient thermoelectric devices.

In this Rapid Communication we consider a molecular quantum dot with an {\em attractive} 
onsite Coulomb interaction, $U<0$, described by a negative-$U$ Anderson impurity model, 
Eq.~(\ref{model}) below. 
Such a model has been used to explain the dielectric properties of
amorphous semiconductors \cite{anderson.75}, to describe highly polarized heavy
fermion states \cite{taraphder.91} and to investigate 
the noise and non-equilibrium transport through negative-$U$ molecules \cite{koch.07}. 
For a molecular quantum dot, several mechanisms could result in $U<0$, for example,
screening by electrons in metallic leads can reduce an initially repulsive 
local Coulomb interaction to negative values \cite{lcuo}, or, 
a vibrating molecule with a local electron-phonon interaction could result
in a net attractive Coulomb interaction \cite{cornaglia.04,hewson.04}. 
For typically used metallic electrodes, such as gold, screening is expected 
to ensure the locality of the attractive interaction in Eq.~(\ref{model}).

A negative-$U$ quantum dot supports a charge Kondo effect in which 
the role of spin-up and spin-down states in the conventional spin Kondo effect 
are played by the non-magnetic empty and doubly occupied states of the dot 
\cite{taraphder.91}. 
As in the usual spin Kondo effect, this 
charge Kondo effect results in a renormalized Fermi liquid at low temperatures
which has important consequences for electrical and thermal transport.
It is also believed to be the origin of superconductivity in PbTe doped with Tl,
where the valence skipper Tl acts as a negative-$U$ center 
\cite{dzero.05,matsushita.05}. 
While some aspects of the electrical transport through a negative-$U$ 
molecule have been investigated \cite{koch.07}, 
the most interesting feature of such a system, elucidated below, 
lies in its remarkable low-temperature Kondo-induced thermoelectric response 
which, to the best of our knowledge, has not been previously addressed. 

{\em Model and calculations.---} 
Specifically, we consider a quantum dot described by the 
following two-lead Anderson impurity model
\begin{eqnarray}
H &=& \sum_{\sigma}\varepsilon_{d}n_{d\sigma} + U n_{d\uparrow}n_{d\downarrow}
+\sum_{k\alpha\sigma}\epsilon_{k\alpha}c_{k\alpha\sigma}^{\dagger}c_{k\alpha\sigma}
\nonumber\\
&+& \sum_{k\alpha\sigma}(t_{\alpha}c_{k\alpha\sigma}^{\dagger}d_{\sigma}+ h.c.),\label{model}
\end{eqnarray}
where, $\varepsilon_{d}$ is the energy of the molecular level, 
$U<0$ is the local Coulomb interaction, $\sigma$  labels the spin, and  
$\alpha=L,R$ labels left and right electron lead states with kinetic 
energies $\epsilon_{k\alpha}$. The couplings of the dot to the leads  are
denoted by $\Gamma_{\alpha}(\omega)=
2\pi \rho_{\alpha}(\omega)|t_{\alpha}|^{2}$, 
where $\rho_{\alpha}(\omega)=\sum_{k}\delta(\omega-\epsilon_{k\alpha})$ is
the density of states of lead $\alpha$.

The linear response transport properties can be calculated from the 
single-particle spectral function of the dot
$A_{\sigma}(\omega)
=
-{\rm Im}[G_{d\sigma}(\omega+i\delta)]/\pi,$
where 
$G_{d\sigma}(\omega+i\delta)=\langle\langle d_{\sigma}; 
d_{\sigma}^{\dagger}\rangle\rangle$
is the Fourier transform of the retarded single-particle Green function of (\ref{model}). 
The thermopower is given by \cite{kim.02}
\begin{eqnarray}
S &=& -\frac{1}{|e|T}\frac{\int d\omega\; \omega{\cal T}(\omega)
\left(-\partial f/\partial \omega\right)}{
\int d\omega {\cal T}(\omega)
\left(-\partial f/\partial\omega\right)},\label{thermo}
\end{eqnarray}
where $f$ is the Fermi function, $e$ is the electronic charge, 
and ${\cal T(\omega)}=2\pi \Gamma(\omega)\sum_{\sigma}A_{\sigma}(\omega)$ 
is the transmission
function of the dot with $\Gamma(\omega) = \frac{\Gamma_{L}(\omega)\Gamma_{R}(\omega)}{
\Gamma_{L}(\omega)+\Gamma_{R}(\omega)}$. At low temperature, a Sommerfeld
expansion leads to
\begin{equation}
S(T)=-\frac{\pi^{2}k_{B}}{3|e|}k_{B}T
\left(\frac{\Gamma'(\epsilon_{F})}{\Gamma(\epsilon_{F})} 
+\frac{\sum_{\sigma}A_{\sigma}'(\epsilon_{F})}{\sum_{\sigma}A_{\sigma}
(\epsilon_{F})}\right)\label{sommerfeld}
\end{equation}
where $\epsilon_{F}=0$ is the Fermi level of the leads. In the absence of 
a magnetic field $A_{\uparrow}(\omega)=A_{\downarrow}(\omega)=A(\omega)$ 
is spin independent. A large thermopower
at low temperature can be achieved by either tailoring the band structure of
the leads to give a highly asymmetric $\Gamma(\omega)$ at $\epsilon_{F}$ with a
large slope $\Gamma'(\epsilon_{F})$ (Ref.~\onlinecite{hicks.93}) or
tailoring correlations to yield a highly asymmetric $A(\omega)$
at $\epsilon_{F}$ with a large slope $A'(\epsilon_{F})$, or both.
We concentrate on the latter which is robust to details of the
lead density of states, and assume a smooth $\Gamma(\omega)$ around $\epsilon_{F}$, 
i.e., we take $\Gamma(\omega)=\Gamma=0.01$ (in units of the half bandwidth of the leads).

The frequency and temperature dependence of  
$A(\omega,T)$ is calculated by using the numerical 
renormalization group (NRG) method \cite{nrg}. 
Results for $U/\Gamma=-8$ were obtained at gate voltages 
$-|e|V_{g}=(\varepsilon_{d}+U/2)$ 
in the range $|V_{g}|\le 8\Gamma$ (setting $e=1$). In addition, for $T=0$, 
we have compared results for occupation numbers $n_{d}$ with those from 
functional renormalization group (fRG)\cite{fRG} and Bethe ansatz \cite{ba} 
techniques (see Fig.~\ref{fig3} below). In the following, 
$T_{K}=\sqrt{|U|\Gamma/4}e^{-\pi |U|/4\Gamma}$ (Ref.~\onlinecite{hewson.97}) 
denotes the relevant
low energy charge Kondo scale of (\ref{model}). Due to the exponential dependence on 
$U$ and $\Gamma$, $T_{K}$ can vary by orders of magnitude, 
e.g. for positive-$U$ systems from $1$\, 
to $200$\, {\rm K} \cite{parks.10}. For $U=-8\Gamma$, 
we have $T_{K}=2.64\times 10^{-3}\Gamma \ll \Gamma$. 

\begin{figure}
\includegraphics[width=\linewidth,clip]{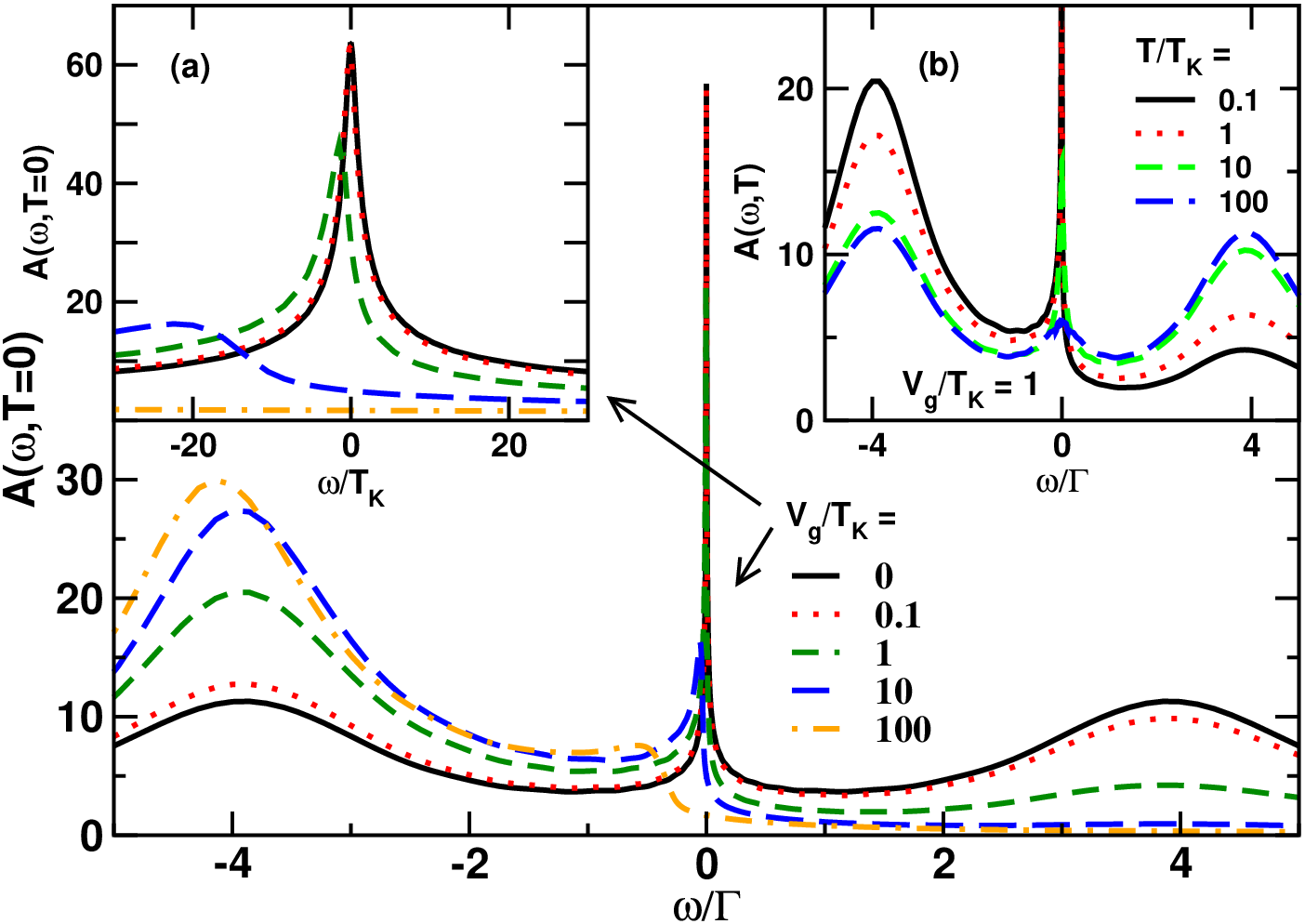}
\caption{{\em (Color online)} Main panel: $T=0$ spectral function for $U/\Gamma=-8$ for 
different gate voltages $V_{g}/T_{K}$. Inset (a): $A(\omega,T=0)$ near $\omega=0$. 
Inset (b): Temperature dependence of $A(\omega,T)$ for $V_{g}/T_{K}=1$.
\label{fig1}}
\end{figure}
\begin{figure}
\includegraphics[width=\linewidth,clip]{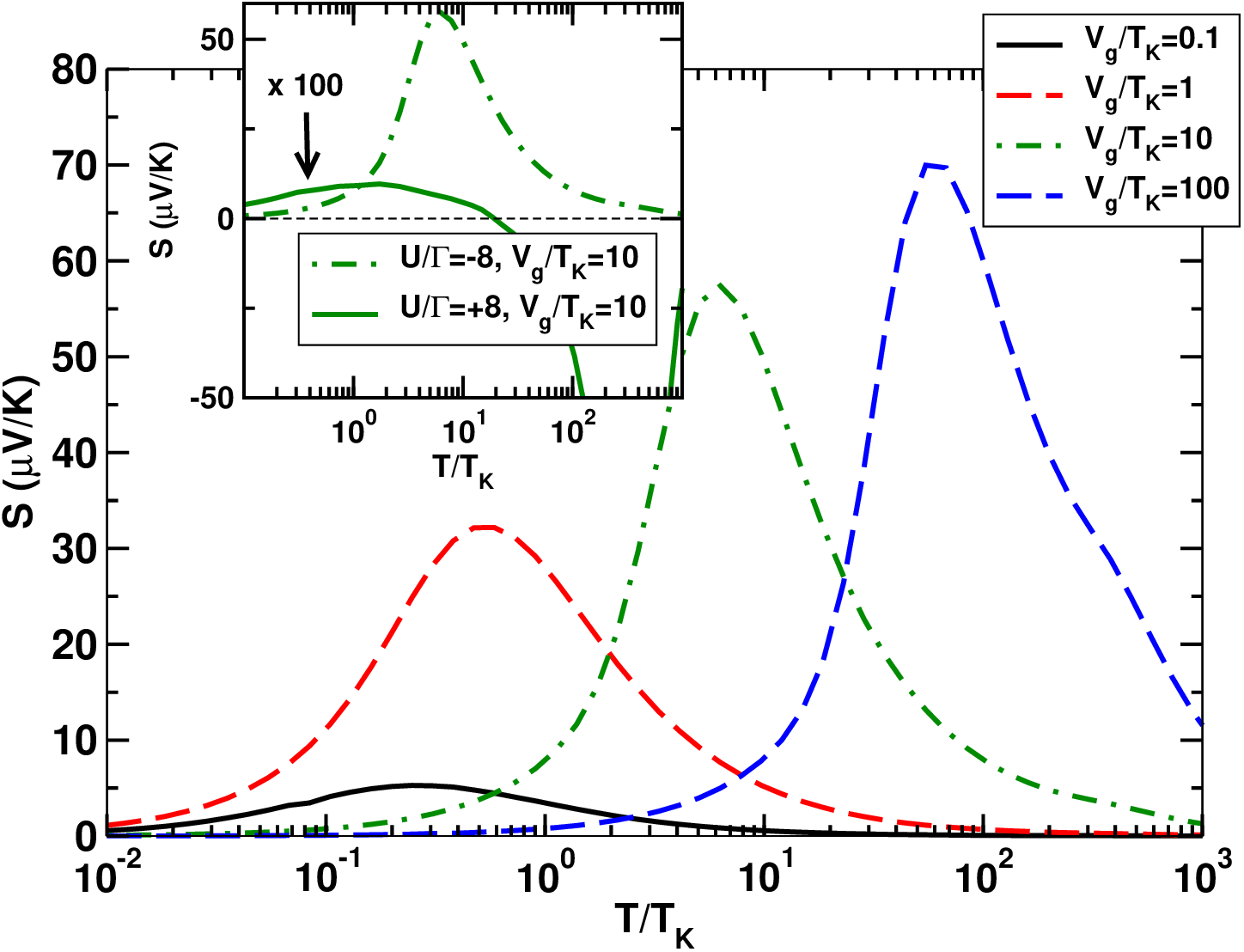}
\caption{{\em (Color online)} Thermopower $S$ vs temperature at different
gate voltages $V_{g}/T_{K}$ and $U/\Gamma=-8$. Inset: Comparison with $U>0$ thermopower
for $V_{g}/T_{K}=10$.
\label{fig2}}
\end{figure}
{\em Results.---}
Figure~\ref{fig1} shows the $T=0$ spectral function for several gate voltages.
At $V_{g}=0$ the pseudo-spin states $n_{d}=0$ and $n_{d}=2$ are degenerate, 
and the spectral function is symmetric, 
with a Kondo resonance of width ${\cal{O}}(T_{K})$
at $\omega=0$ and two Hubbard satellite peaks at 
$\omega=\varepsilon_{d}>0$ and $\omega=\varepsilon_{d}+U <0$.
A finite gate voltage $V_{g}$ induces a splitting $\Delta E = -2V_{g}$ of 
the pseudo-spin states which is analogous to a magnetic field in the 
conventional spin Kondo effect,
i.e. the spectral function becomes highly asymmetric due to the polarizing effect
of $V_{g}$, with $n_{d}$ changing substantially from its ``perfectly 
screened'' value of $n_{d}=1$ \cite{spectral-asymmetry}. 
This asymmetry in the single-particle 
spectral function with a large slope at $\epsilon_{F}$, 
for {\em both} spin components, is the origin of the large thermopowers 
to be discussed below. The analogy to the spin Kondo effect in a magnetic
field, can be made precise for the case of particle-hole symmetric bands which we consider: 
A particle-hole transformation on the down spins 
allows the negative-$U$ Anderson model in the absence of a local 
magnetic field to be mapped onto
the positive-$U$ symmetric Anderson model in a finite local magnetic field 
$B=2\epsilon_{d}+|U|=-2V_{g}$ \cite{iche.72}, thereby explaining 
the highly asymmetric spectral function of (\ref{model}) shown in Fig.~\ref{fig1}. 
The polarizing effect of finite $V_{g}\sim B$ is strongest at 
$T=0$ and diminishes for $T\gg T_{K}$ [see Fig.~\ref{fig1}(b)]. In terms of the above
analogy, this corresponds to the quenching of the magnetization 
$M=(n_{d\uparrow}-n_{d\downarrow})/2$ at high temperatures in the 
corresponding positive-$U$ model in a field $B$. 

Figure~\ref{fig2} shows the main result of this Rapid Communication: 
a dramatic enhancement of the
Seebeck coefficient induced by a finite gate voltage $V_{g}\gtrsim T_{K}$ exceeding
$50\mu V/K$ for $V_{g}\gtrsim 2T_{K}$. The maximum in the thermopower occurs on a 
temperature scale which correlates with $V_{g}$ and is therefore highly tunable. 
Corresponding Seebeck coefficients for $U>0$ in the Kondo regime are 
insignificant (see the inset of Fig.~\ref{fig2}).
The large enhancement in $S$ is due to the correlation-induced asymmetry 
in the spectral function at finite $V_{g}$. At low temperatures, explicit
calculations, within Fermi-liquid theory \cite{hewson.97}, also shed light
on this enhancement. In this limit, the thermopower may be expressed in terms 
of the occupancy $n_{d}$ of the dot as
\begin{equation}
\label{thermo-fliq}
S(T) =-\frac{\pi\gamma T}{|e|}\cot(\pi n_{d}/2)
\end{equation} 
with $\gamma T \ll 1$, and $\gamma$ being the linear coefficient of specific heat 
of the dot \cite{hewson.97}
(with $\gamma\sim 1/T_{K}$ for $V_{g}=0$). 
A finite $V_{g}\sim T_{K}$ polarizes the charge Kondo state,
leading to $n_{d}\sim 2$ for $V_{g}>0$. This enhances a nominally small ($\ll k_{B}/|e|$) 
thermopower by the large factor $\cot(\pi n_{d}/2)\gg 1$. Note also,
that while a finite magnetic
field for $U>0$ also leads to asymmetric spectral functions $A_{\uparrow}(\omega)$ and 
$A_{\downarrow}(\omega)$ around $\epsilon_{F}$, the asymmetry in the Kondo regime is opposite
for spin up and spin down. Consequently, it largely cancels in the combination 
$\sum_{\sigma}A_{\sigma}$ entering (\ref{thermo}) and the thermopower is not enhanced. 
[Furthermore, for $n_d\simeq 1$, the factor $\cot(\pi n_{d}/2)$ is very small.]
Finally, note that $S(T)$ of the negative-$U$ model in Fig.~{\ref{fig2}} does not 
exhibit a sign change with increasing temperature for any finite $V_{g}$, 
in contrast to the case of $U>0$ \cite{costi.10}: The sign
change of the latter is due to a change in slope of the spectral function at the
Fermi level, induced by a collapse of the Kondo resonance with increasing temperature. 
This cannot occur in the $U<0$ model, since the spectral function remains 
polarized by a finite $V_{g}$ at all relevant temperatures. 

\begin{figure}
\includegraphics[width=\linewidth,clip]{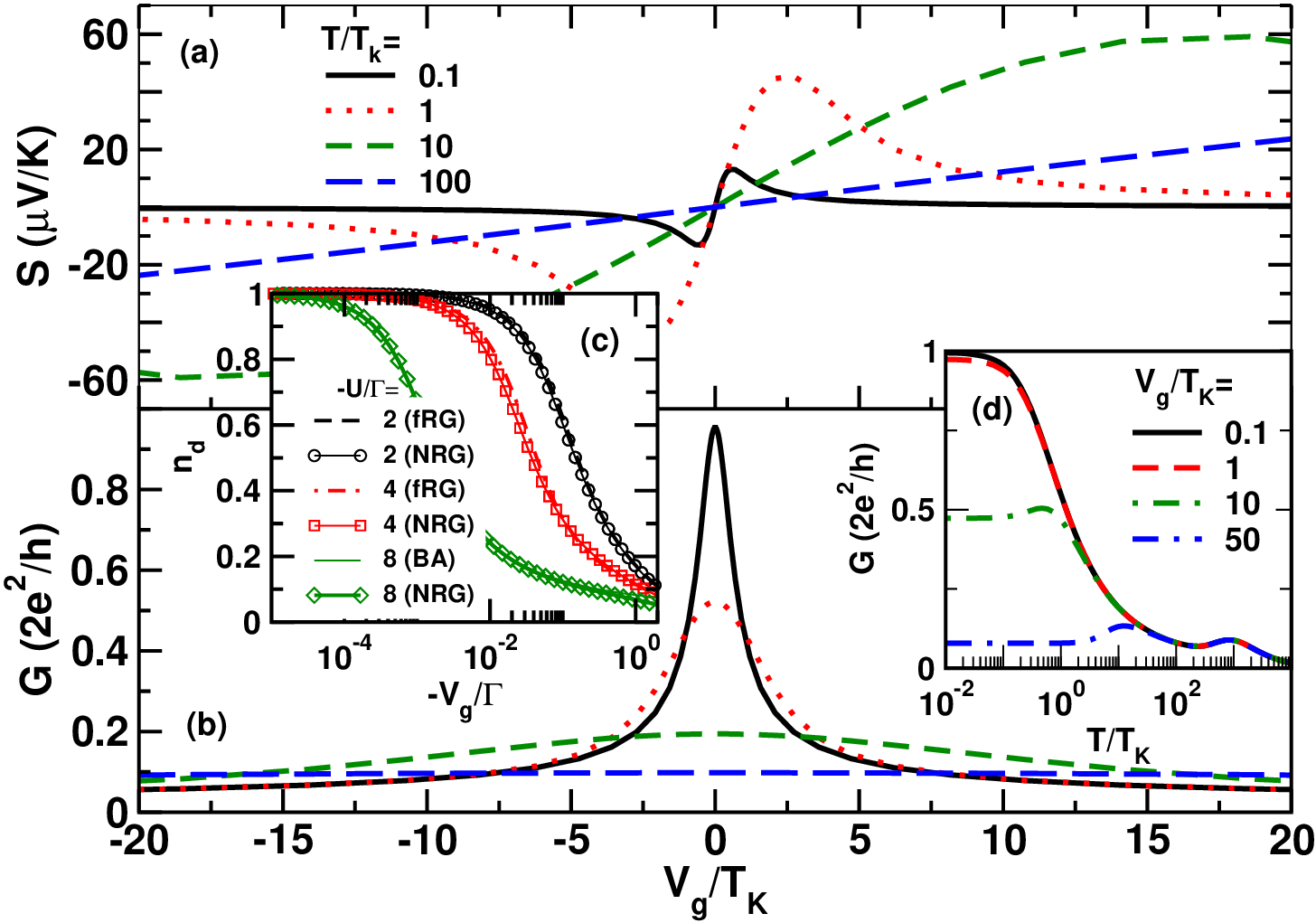}
\caption{{\em (Color online)} Gate voltage dependence of thermopower $S$ (a) 
and conductance $G$ (b) at typical temperatures $T/T_{K}$. Inset (c): Dot 
occupation number $n_{d}$ vs gate voltage for $U/\Gamma=-2,-4,-8$. fRG results
for $U/\Gamma=-2,-4$ agree with NRG to less than 8\% relative error, while for 
$U/\Gamma=-8$ NRG agrees with the Bethe ansatz (Ref.~\onlinecite{ba}) very well. 
Inset (d): Temperature dependence of $G$ at selected gate voltages $V_{g}/T_{K}$.
  \label{fig3}}
\end{figure}

The gate voltage dependence of the thermopower and electrical
conductance is shown in Fig.~\ref{fig3}(a) and 3(b) at several temperatures. 
Except at $T\lesssim T_{K}$, a large Seebeck coefficient exceeding 
$50\mu V/K$ can always be realized by a suitable choice of gate voltage. By
tuning the gate voltage to positive or negative values about the charge
Kondo state at $V_{g}=0$ one can realize the $p$-type or $n$-type legs of a
thermoelectric device.
Note the absence of a Kondo plateau in $G(V_{g})$ at $T\ll T_{K}$ in Fig.~\ref{fig3}(b), 
which contrasts with the $U>0$ case, and the rapid drop on a scale $V_{g}\sim T_{K}$ of
$G(V_{g})$ due to the suppression of the Kondo state by the 
finite gate voltage acting as a magnetic field in the conventional Kondo effect 
\cite{koch.07,cornaglia.04,wysokinski.08}. 
NRG results for $n_{d}(T=0)$ vs $V_{g}$ compare very well with 
fRG calculations at $U/\Gamma =-2, -4$ and with exact Bethe Ansatz (BA) 
calculations at $U/\Gamma = -8$ [see Fig.~\ref{fig3}(c)]. Figure~\ref{fig3}(d)
shows that $G(T)$ exhibits the typical Kondo scaling behavior at small gate voltages 
$V_{g}\lesssim T_{K}$.

\begin{figure}
\includegraphics[width=\linewidth,clip]{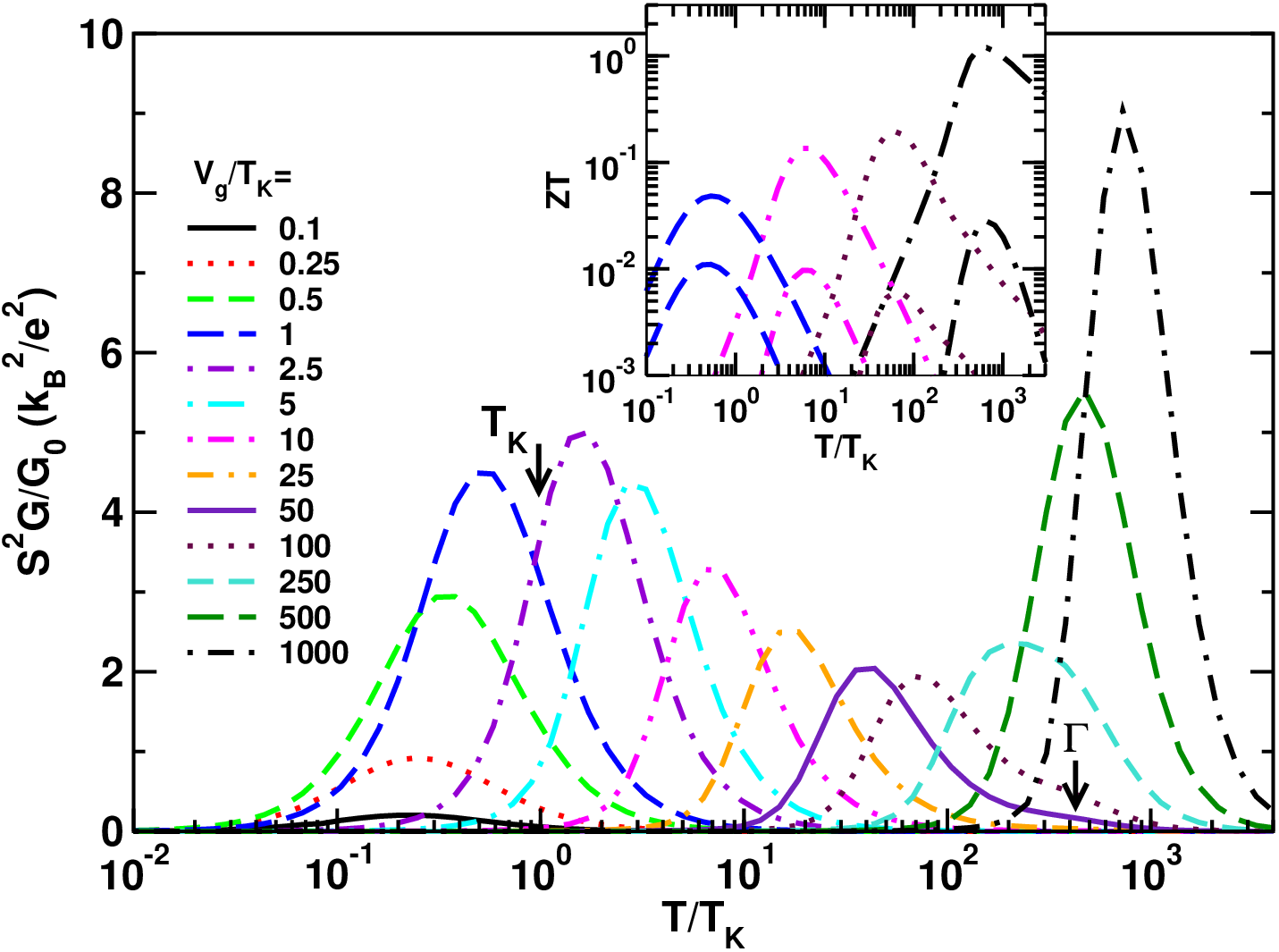}
\caption{{\em (Color online)} 
The power factor $S^{2}G$ divided by $G_{0}=2e^{2}/h$ vs temperature and for
a range of gate voltages $V_{g}/T_{K}$ for $U/\Gamma=-8$. The locations of $T_{K}$ and
$\Gamma$ are indicated by arrows. Inset: Upper and lower bounds for $ZT$, as defined
in the text, at several gate voltages. 
  \label{fig4}}
\end{figure}
The thermoelectric efficiency of a nanoscale device is related to its 
dimensionless figure of merit defined by $ZT=PT/K$, where $P=S^{2}G$
is the power factor, and $K= K_{\rm e}+K_{\rm ph}$ the thermal conductance due 
to electrons (e) and phonons (ph). For metallic leads, 
$K_{\rm e}$ will give the dominant contribution to $K$ \cite{segal.03},
while for semiconducting leads, $K_{\rm ph}$ will also be important. Since, to the
best of our knowledge, no calculation of $K_{\rm ph}$ in the presence of Kondo correlations
is available, we discuss the efficiency of our system in terms of the power 
factor $P_{V_g}(T)$, shown in Fig.~\ref{fig4}, and give upper and lower 
bound estimates for $ZT$ below.  The power factor is largely independent 
of details of the leads, making it a useful quantity for future comparison 
with experiments. It is also a relevant quantity for on-chip cooling of a
hot source in microelectronics \cite{mahan.97}. 
For each $V_{g}$ the power factor exhibits a maximum at a temperature which is
related to $V_{g}$. The envelope of these curves has two maxima, one
at $T\approx T_{K}$ for $V_{g}\approx 2T_{K}$ and another at high temperatures 
$T\approx 2\Gamma$ for $V_{g}\sim 4\Gamma \gg T_{K}$. 
In contrast, for $U>0$, the power factor is vanishingly small in the Kondo regime, 
with larger values being obtained only at the border between mixed valence 
and Kondo regimes \cite{costi.10}. Turning to $ZT$, an upper bound estimate 
is obtained by setting $K_{\rm ph}=0$.
A lower bound estimate is obtained by assuming that the molecule is transparent to 
phonons. In this case, each phonon mode contributes the maximum ballistic 
thermal conductance of $\kappa_{0}=\pi^{2}k_{B}T/3h$ \cite{rego.98}. For
three phonon modes we have $K_{\rm ph}=3\kappa_{0}$ resulting in a lower bound 
estimate for $ZT$. Both bounds (see the inset of Fig.~\ref{fig4})
show a maximum at a temperature $T$ that correlates with $V_{g}$, 
with the upper bound exceeding $1$ for $V_{g}/T_{K}\gg 1$ and $T/T_{K}\gg 1$. 
In a real device, phonons will be inelastically scattered, e.g. by
vibrational modes of the molecule, thereby reducing $K_{\rm ph}$ below 
its ballistic value, especially at higher temperatures where anharmonic 
effects become important. Hence, our lower bound for $ZT$ is likely too 
stringent so that a suitable choice of gate voltage could allow 
interesting values of $ZT\sim 0.5-1$ to be achieved at $T\sim 100 T_K$.

{\em Conclusions.---} 
In summary, we investigated the thermoelectric properties of 
a negative-$U$ molecular quantum dot exhibiting the charge Kondo effect. 
A small gate voltage $V_{g}\gtrsim T_{K}$ is found to polarize the
charge on the dot creating a single-particle spectral function 
which is highly asymmetric about the Fermi level. This yields 
a large enhancement of the Seebeck coefficient 
exceeding $50\mu V/K$ on a temperature scale comparable to $V_{g}$. 
The device is highly tunable and allows large power factors to be achieved
at virtually any temperature by a suitable choice of gate voltage. 
In addition to the above-mentioned possible realizations of such devices,
molecular complexes similar to those in Ref.~\onlinecite{parks.10}, but 
with valence skipping ions \cite{varma.88} such as Bi, Tl or In, 
acting as negative-$U$ centers, and attached to gold leads, could be 
promising systems to look into in the future. Reducing the 
dimensionality of the leads, e.g. by using carbon nanotubes 
\cite{guo.07}, could further enhance the power factor \cite{hicks.93}.
\acknowledgments We acknowledge supercomputer support by the John von
Neumann Institute for Computing (J\"ulich), discussions with
C. Karrasch and V. Meden (S.A.), and support by NSF Grant DMR-1006605,
Forschungszentrum J\"{u}lich and the Ministry of Science, Croatia (V.Z.).


\vspace{-0.5cm}

\begin{thebibliography}{10}

\bibitem{mahan.97} 
  G. D. Mahan, 
  Solid State Phys. {\bf 51}, 82 (1997).

\bibitem{snyder.08} 
  G. J. Snyder and E. S. Toberer, 
  Nature Mater. {\bf 7}, 105 (2009).

\bibitem{kanatzidis.10} 
  M. G. Kanatzidis, 
  Chem. Mater. {\bf 22}, 648 (2010).

\bibitem{reddy.07} 
  P. Reddy, S-Y. Jang, R. A. Segalman, and A. Majumdar, 
  Science {\bf 315}, 1568 (2007).

\bibitem{hewson.97} 
  A. C. Hewson, 
  \emph{The Kondo Problem To Heavy Fermions,} Cambridge Studies in Magnetism 
  (Cambridge University Press, Cambridge, U. K., 1997).

\bibitem{glazman.88} 
  L. I. Glazman and M. E. Raikh, 
  JETP Lett. {\bf 47}, 452 (1988).

\bibitem{ng.88} 
  T.-K. Ng and P. A. Lee, 
  Phys. Rev. Lett. {\bf 61}, 1768 (1988).

\bibitem{goldhaber-gordon.98} 
  D. Goldhaber-Gordon, J. G\"{o}res, M. A. Kastner, Hadas Shtrikman, D. Mahalu, and U. Meirav,
  Phys. Rev. Lett. {\bf 81}, 5225 (1998).

\bibitem{cronenwett.98} 
  S. M. Cronenwett, T. H. Osterkamp, and L. P. Kouwenhoven, 
  Science {\bf 281}, 540 (1998).

\bibitem{vanderwiel.00} 
  W. van der Wiel, S. De Franceschi, T. Fujisawa, 
  J. M. Elzerman, S. Tarucha, and L. P. Kouwenhoven,
  Science {\bf 289}, 2105 (2000).

\bibitem{scheibner.05} 
  R. Scheibner, H. Buhmann, D. Reuter, M. N. Kiselev, and L. W. Molenkamp,  
  Phys. Rev. Lett. {\bf 95}, 176602 (2005).

\bibitem{kim.02}T-S. 
  Kim and S. Hershfield, 
  Phys. Rev. Lett. {\bf 88}, 136601 (2002).

\bibitem{costi.10} 
  T. A. Costi and V. Zlati\'c, 
  Phys. Rev. B {\bf 81}, 235127 (2010).

\bibitem{anderson.75} 
  P. W. Anderson, 
  Phys. Rev. Lett. {\bf 34}, 953 (1975).

\bibitem{taraphder.91} 
  A. Taraphder and P. Coleman,
  Phys. Rev. Lett. {\bf 66}, 2814 (1991).

\bibitem{koch.07} 
  J. Koch, E. Sela, Y. Oreg, and F. von Oppen,
  Phys. Rev. B {\bf 75}, 195402 (2007).

\bibitem{lcuo} 
  I. E. Perakis and C. M. Varma,
  Phys. Rev. B {\bf 49}, 9041 (1994); 
  T. A. Costi, Phys. Rev. B {\bf 55}, 6670 (1997).

\bibitem{cornaglia.04} 
  P. S. Cornaglia, H. Ness, and D. R. Grempel,
  Phys. Rev. Lett. {\bf 93}, 147201 (2004).

\bibitem{wysokinski.08} For results at $|U|\gg T \gg \Gamma$, 
  see M. Gierczak and K. I. Wysoki\'{n}ski,
  J. Phys.: Conf. Series {\bf 104}, 012005 (2008).

\bibitem{hewson.04}
  A. C. Hewson, A. Oguri, and D. Meyer, 
  Eur. Phys. J. B {\bf 40}, 177 (2004).

\bibitem{dzero.05} 
  M. Dzero and J. Schmalian, 
  Phys. Rev. Lett. {\bf 94}, 157003 (2005).

\bibitem{matsushita.05} Y. Matsushita, H. Bluhm, T. H. Geballe, and I. R. Fisher,
  Phys. Rev. Lett. {\bf 94}, 157002 (2005).

\bibitem{hicks.93} L. D. Hicks and M. S. Dresselhaus, 
  Phys. Rev. B {\bf 47}, 12727 (1993).

\bibitem{nrg} K. G. Wilson, 
  Rev. Mod. Phys. {\bf 47}, 773 (1975); 
  R. Bulla, T. A. Costi, and T. Pruschke, 
  {\em ibid.} {\bf 80}, 395 (2008); 
  W. Hofstetter,
  Phys. Rev. Lett. {\bf 85}, 1508 (2000);
  R. Peters, T. Pruschke, and F. B. Anders,
  Phys. Rev. B {\bf 74}, 245114 (2006);
  A. Weichselbaum and J. von Delft,
  Phys. Rev. Lett. {\bf 99}, 076402 (2007).

\bibitem{fRG} M. Salmhofer and C. Honerkamp, 
  Prog. Theor. Phys. {\bf 105}, 1 (2001); 
  C. Karrasch, T. Enss, and V. Meden, 
  Phys. Rev. B {\bf 73}, 235337 (2006).

\bibitem{ba} A. M. Tsvelick and P. B. Wiegmann, 
  Advances in Physics {\bf 32}, 453 (1983); 
  N. Andrei, Phys. Lett. {\bf 87}A, 299 (1982).

\bibitem{parks.10} J. J. Parks, A. R. Champagne, T. A. Costi, 
  W. W. Shum, A. N. Pasupathy,
  E. Neuscamman, S. Flores-Torres, P. S. Cornaglia, 
  A. A. Aligia, C. A. Balseiro, G. K.-L. Chan,
  H. D. Abru\~{n}a, and D. C. Ralph,
  Science {\bf 328}, 1370 (2010).

\bibitem{spectral-asymmetry} A. Rosch, T. A. Costi, 
  J. Paaske, and P. W\"{o}lfle,
  Phys. Rev. B {\bf 68}, 014430 (2003).

\bibitem{iche.72} G. Iche and A. Zawadowski, 
  Solid State Commun. {\bf 10}, 1001 (1972); 
  A. C. Hewson, J. Bauer, and W. Koller,
  Phys. Rev. B {\bf 73}, 045117 (2006). 

\bibitem{segal.03} D. Segal, A. Nitzan, and P. H\"{a}nggi, 
  J. Chem. Phys. {\bf 119}, 6840 (2003).

\bibitem{rego.98} L. G. C. Rego and G. Kirczenow, 
  Phys. Rev. Lett. {\bf 81}, 232 (1998).

\bibitem{varma.88} C. M. Varma, 
  Phys. Rev. Lett. {\bf 61}, 2713 (1988).

\bibitem{guo.07} X. Guo, A. Whalley, J. E. Klare, L. Huang, 
  S. O'Brien, M. Steigerwald, and C. Nuckolls,
  Nano Letters {\bf 7}, 1119 (2007).

\end{thebibliography}
\end{document}